\pdfoutput=1

\documentclass[preprint,showpacs,preprintnumbers,amsmath,amssymb]{revtex4}

\usepackage{graphicx}
\usepackage{dcolumn}
\usepackage{bm}

\begin{document}

\preprint{UCD-0905}

\title{Computation of the string tension in 
four-dimensional Yang-Mills theory using large $N$ reduction}

\author{Joe Kiskis}
\affiliation{Department of Physics, University of California, Davis, 
CA 95616, USA}
\email{jekiskis@ucdavis.edu}
\author{Rajamani Narayanan}
\affiliation{Department of Physics, Florida International University, 
Miami, FL 33199, USA}
\email{rajamani.narayanan@fiu.edu}

\date{\today}


\begin{abstract}
Continuum reduction and Monte Carlo simulation are used to calculate the 
heavy quark potential and the string tension in large $N$ Yang-Mills 
theory in four dimensions. The potential is calculated out to a 
separation of nine lattice units on a lattice with extent six in each 
direction. 
\end{abstract}


\maketitle

\section{Introduction}

Through dimensional transmutation, the dimensionless classical coupling 
constant of QCD is transformed to a running coupling controlled by a 
physical scale $\Lambda_{QCD}$. Roughly speaking, weak coupling 
perturbative approximations to processes with momentum transfer $Q$ are 
expansions in $1/ln(Q/\Lambda_{QCD}$). This gives excellent results when 
$Q/\Lambda_{QCD}$ is large. When one cannot rely on that, an 
alternative is the large $N$ approach. 
The gauge group is generalized to $SU(N)$, and the expansion is in 
$1/N$. 

Although the large $N$ limit of Yang-Mills theory is still out of reach, 
it is known that it enjoys several simplifications. Among them is the 
possibility of reducing the space-time volume without affecting certain 
physical results~\cite{ek}. Continuum reduction~\cite{nn1,knn2}, {\it 
i.e.} reduction to a 
physical size of order $1/\Lambda_{QCD}$ avoids some of the 
difficulties~\cite{bhn,bs,tv} 
with reduction to a single space-time point.
The addition of double trace terms to the action~\cite{unsal} is an 
alternative approach.

It is now practical to obtain good results by combining continuum 
reduction 
with numerical simulations needing only modest resources. In previous 
numerical work~\cite{nn2,nn3,kn3d,kn3ds},
it has been shown that not only bulk quantities but also 
physical quantities based on Wilson loops are accessible. The previous 
results based on Wilson loops have been in three dimensions. In the work 
described here, we show that the method is still practical in four 
dimensions.

We have used continuum reduction and Monte Carlo simulation to calculate 
the heavy quark potential and the string tension in large $N$ Yang-Mills 
theory in four dimensions. An important aspect of the method is that 
reduction allows the calculation of infinite volume, infinite $N$ Wilson 
loops that are larger than the reduced lattice. In particular, in this 
work, the heavy quark potential is calculated out to a 
separation of nine lattice units on a lattice with extent six in each 
direction. The results for the string tension are compatible with 
those obtained on large lattices at smaller $N$.

\section{Methods}

The standard Wilson Yang-Mills action with gauge group $SU(N)$ is used. 
In the large $N$ limit, $g^2$ is taken to zero with the inverse 't Hooft 
coupling $b=\frac{1}{g^2N}$ held fixed.

We report results with $N=37$, $47$, and $59$
on a lattice of size $6^4$. This is 
large enough so that in the range of available $b$, the lattice is not 
too coarse but is still small enough to give a manageable computational 
cost.

In four dimensions, the useful range of couplings and physical lattice 
spacings $a$ on a given lattice is more limited than in three 
dimensions. For $6^4$, the system becomes unstable to the bulk 
transition for small $b$. The smallest $b$ we have used is 0.3450, which 
is just above the unstable point. For sufficiently large $b$, the 
center symmetry breaks in one lattice direction, and reduction no longer 
holds in that direction. This is the large $N$ limit of the finite 
temperature phase transition~\cite{knn2,jk} . For $6^4$ it occurs at 
about 
$b=0.3515$. 
Our calculations are at $b=0.3450$, $0.3480$, and $0.3500$ with most of 
the results at $0.3480$. As measured by the critical size for the 
finite-temperature transition, that is a range from about $L_c=4.4$ to 
$5.6$. Thus the possibilities for testing scaling are quite limited. For 
this lattice size, this entire useful range of $b$ is in the region that 
is metastable to the bulk transition. Nevertheless, we encountered no 
bulk transitions during the simulations.

As described in previous work~\cite{knn2}, we update the gauge field 
configurations with heat bath and over-relaxation methods. In this work, 
one update will mean one heat bath sweep followed by one over-relaxation 
sweep. Measurements of the Wilson loops were separated by ten 
updates. The values for the Wilson loops are based on 1200 
measurements.

The measurements of Wilson loops are made on smeared configurations. The 
use of smeared links improves the measurement of Wilson loops. They 
enhance the overlap of the space-like sides of the Wilson loops with the 
ground state. This increases the signal relative to the fluctuations and 
simplifies the $t$ behavior of the loops~\cite{teper_sm}. The 
smearing is a four-dimensional version of the method used 
in~\cite{kn3d}. 
One lattice direction is arbitrarily chosen as the ``time" 
direction. Links in the remaining three spatial directions (but not in 
the time direction) are smeared. After the Wilson loops in these 
``time"-space planes are measured, the process is repeated with each of 
the other lattice directions chosen as ``time." 

When smearing the links in spatial 
directions, only staples in spatial planes are used. One step in the 
iteration takes one from a set $U^{(i)}_k (x_1,x_2,t)$ to a set 
$U^{(i+1)}_k (x_1,x_2,t)$. Before reunitarization, the weight of 
$U^{(i)}_k (x_1,x_2,t)$ is $(1-f)$ while that of each staple is $f/4$. 
There are two parameters, namely, the smearing factor $f$ and the number 
of smearing steps $n$. We use $f=0.45$ and $n=5$ so that $\tau=fn=2.25$ 
and the associated length scale is $\sqrt{\tau}=1.5$. The use of finer 
smearing steps $f=0.1$, $n=25$ or a larger length scale $f=0.45$, $n=10$ 
was more costly and did not lead to further improvement. 

Data were collected on planar, rectangular Wilson loops of size $k 
\times j$ with $k$ and $j$ ranging from 1 to 9 and with $j$ the extent 
in the ``time" direction. The $j$ decay of the loops is fit to a simple 
exponential. To avoid a possible distortion from a combination of 
smearing and very small error bars at the 
shortest separation, loops that are $1 \times j$ and $k \times 1$ are 
not included in the fit.  
The rate of the exponential decay is taken as the static 
quark potential at the separation $k$. The $k$ dependence of the 
exponential is fit to obtain the string tension. With all quantities in 
lattice units, the three parameter potential that is used in the fits is
\begin{equation}
 m(k) = \sigma a^2 k + c_0 + \frac{c_1}{k} \label{pot}
\end{equation} 
We expect $c_0$ to be positive and of order $1/b$ and $c_1$ to be 
negative and of order $1/b$ or O(1).

Errors in all quantities at a fixed $b$ and $N$ are obtained by jack knife 
with single elimination.

\section{Results}

We have results for $N=47$ at $b$ values of 0.3450, 0.3480, and 0.3500. 
In addition, there are results for $N=37$ and $N=59$ at $b=0.3480$.

\begin{figure}
\centering
\includegraphics[scale=0.5]{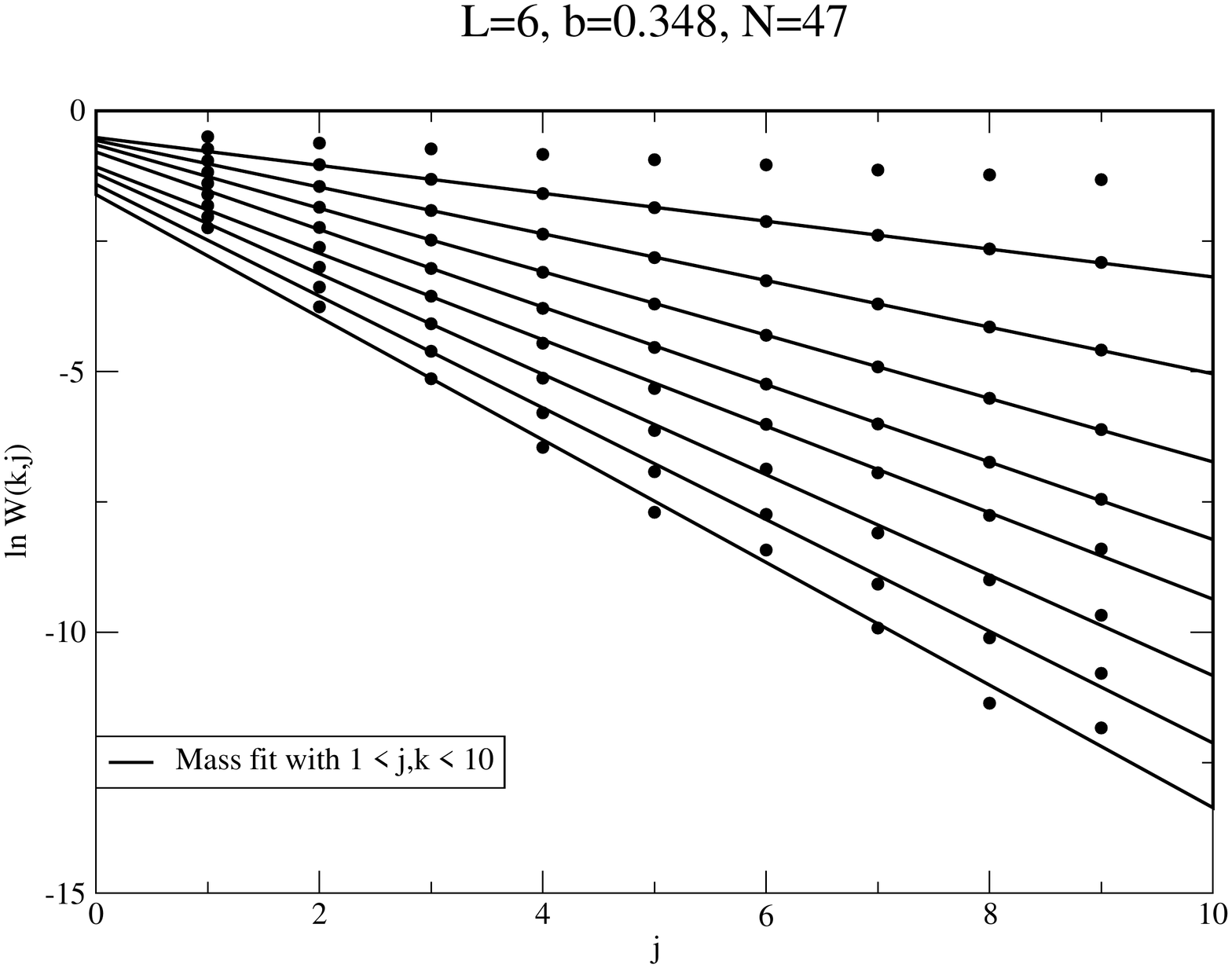}
\caption{\label{fig1}Exponential fits to Wilson loops $W(k,j)$ with $j$ 
on the horizontal axis and $k$ increasing from 1 to 9 going downward.}
\end{figure}

\begin{figure}
\centering
\includegraphics[scale=0.5]{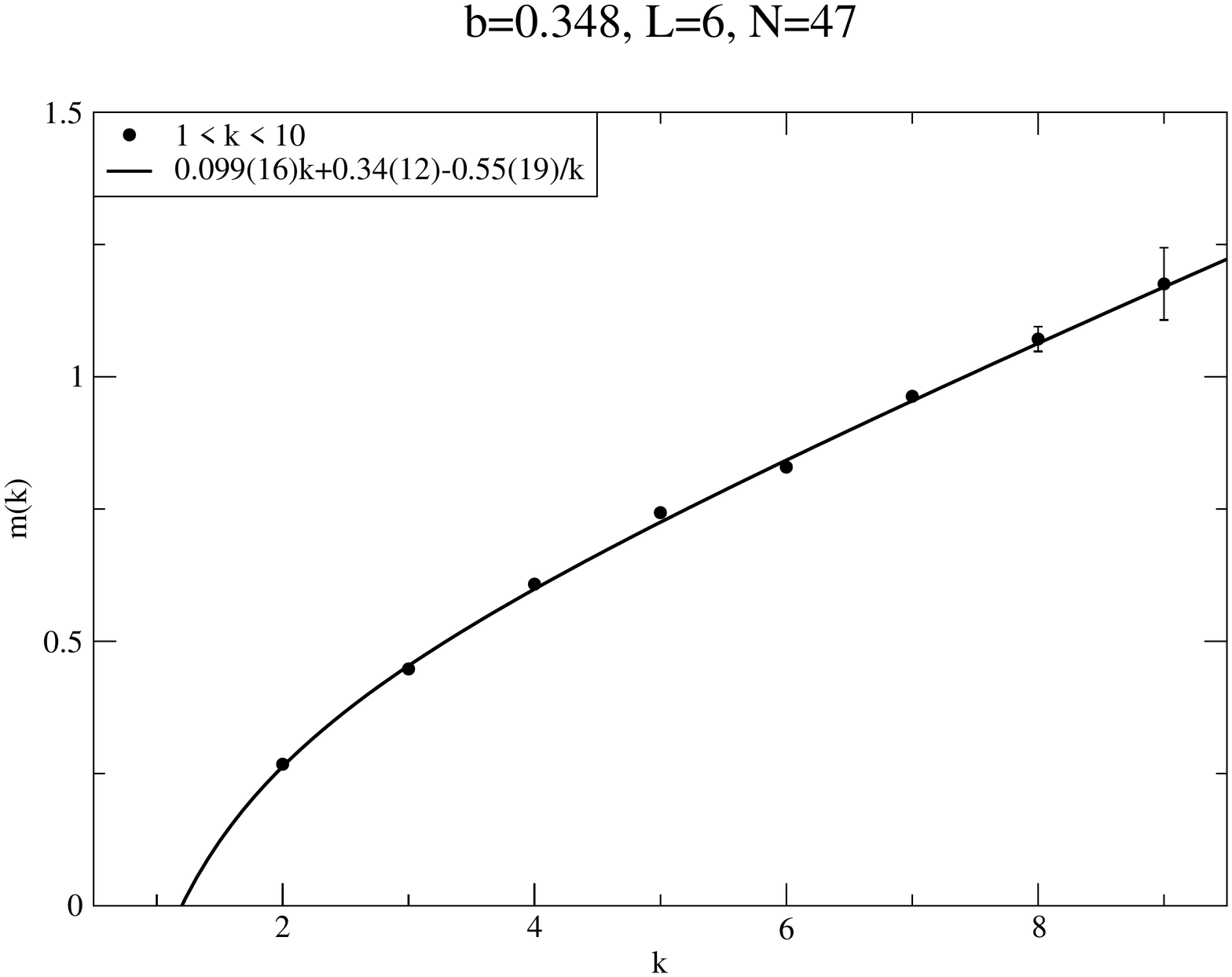}
\caption{\label{fig2}The static quark potential $m(k)$.}
\end{figure}

An example of fits to Wilson loop data is given in Fig.~\ref{fig1}, 
which is for 
$b=0.3480$ and $N=47$.  
At a fixed $k$, the decay in $j$ of loop data is fit to the form 
 $Ae^{-m(k)j}$ using $k$ and $j$ from 2 through 9, inclusive.
This gives the potential $m(k)$ at separation $k$ which is then plotted 
as a function of $k$ in Fig.~\ref{fig2}.
A fit of the potential for the $b=0.3480$ and $N=47$ data to the form 
of Eq.~\ref{pot} gives 
\begin{equation}
  \sigma a^2 = 0.099 \pm 0.016 
\end{equation}

\begin{figure}
\centering
\includegraphics[scale=0.5]{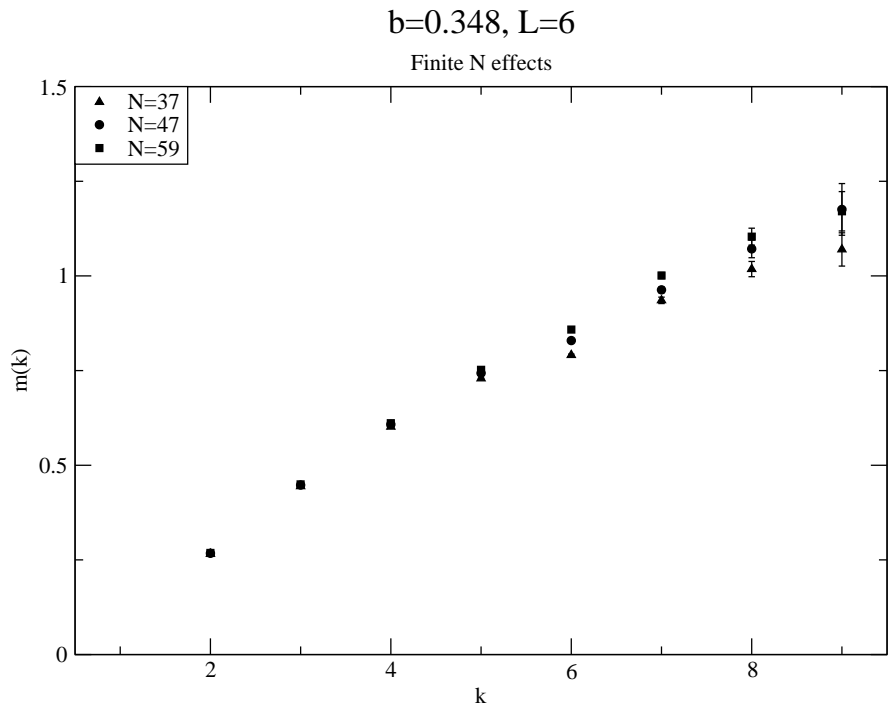}
\caption{\label{fig3} The $N$ dependence of the potential.}
\end{figure}

To verify that $N$ is sufficiently large, we have the results for 
$N=37$, $47$, and $59$ in Fig.~\ref{fig3}.
The string tensions from fits to the $N=47$ and $N=59$ data agree.

\begin{figure}
\centering
\includegraphics[scale=0.5]{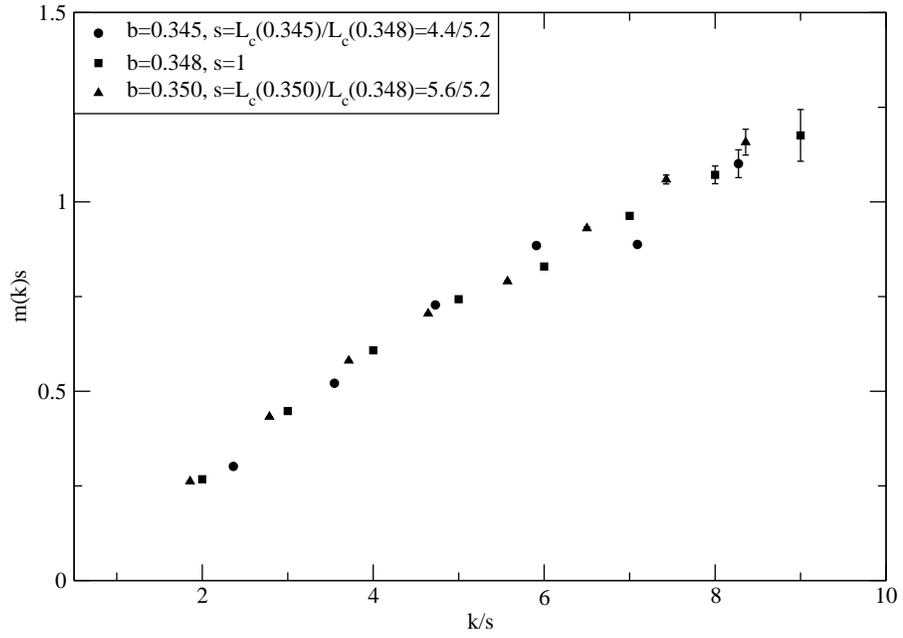}
\caption{\label{fig4}Scaling behavior of the potential.}
\end{figure}

For a check of scaling in the limited range available, we can compare 
the potentials for $b=0.3450$, $0.3480$, and $0.3500$ all with $N=47$.
(For $b=0.3450$, it was necessary to restrict the largest dimension of 
the loops to 8 to obtain useful fits.) 
The relative physical scales are set from ratios of the critical lengths 
$L_c(b)$ at which the center symmetry breaks in one direction. These are 
determined from the results in~\cite{knn2}. With 
\begin{equation}
 s(b) = \frac{L_c(b)}{L_c(0.3480)} ,
\end{equation}
we have $s=4.4/5.2$, $1$, and $5.6/5.2$ for $b=0.3450$, $0.3480$, and 
$0.3500$, respectively. In Fig.~\ref{fig4}, we plot $ms$ verses $k/s$ 
and see that there is rough agreement of the physically scaled 
potentials.

\section{Conclusion}

A comparison with results on large lattices is in order. This can be 
done by comparing either bare quantities at the same tadpole improved 
coupling or by comparing a dimensionless ratio. Our result for the 
string tension in lattice units is $\sigma a^2 = 0.099 \pm 0.016$. This 
is at bare coupling $b = 0.3480$ which corresponds to a tadpole improved 
$b_I = b \langle \Box \rangle = 0.182$. (The average plaquette 
$\langle \Box \rangle$ is normalized to approach one in the weak 
coupling limit.) As it happens, Lucini, Teper, and 
Wenger~\cite{ltw1,ltw2} 
have a large lattice, $N=8$ result $\sigma a^2 = 0.116 \pm 0.001$ at 
the same $b_I$. This falls within our much larger uncertainty range.

An appropriate dimensionless ratio of physical quantities is
$ T_c / \sqrt{\sigma} $ where, in our case, $aT_c = 1/L_c$.
At $b_I = 0.182$, our result is $ T_c / \sqrt{\sigma} = 0.61 \pm 0.05$. 
After extrapolations to infinite volume, infinite $N$, and zero 
lattice spacing, Lucini, Teper, and Wenger~\cite{ltw_1,ltw0,ltw2} obtain 
$0.597 \pm 0.004$.

In conclusion, we have used large $N$   
continuum reduction and Monte Carlo simulation to calculate the
heavy quark potential and the string tension in Yang-Mills
theory in four dimensions. The results are compatible with those 
obtained on large lattices at smaller $N$.

\begin{acknowledgments}

R.N. acknowledges partial support by the NSF under grant number
PHY-0854744.

\end{acknowledgments}



\newpage

\end{document}